# Experimental determination of pore shapes using phase retrieval from *q*-space NMR diffraction

Kerstin Demberg[1,2], Frederik Bernd Laun[3], Marco Bertleff[4], Peter Bachert[1], and Tristan Anselm Kuder[1]*

[1] *Medical Physics in Radiology, German Cancer Research Center (DKFZ), Heidelberg, Germany*

[2] *Faculty of Physics and Astronomy, Heidelberg University, Heidelberg, Germany*

[3] *Institute of Radiology, University Hospital Erlangen, Friedrich-Alexander-Universität Erlangen-Nürnberg, Erlangen, Germany*

[4] *Computer Assisted Clinical Medicine, Medical Faculty Mannheim, Heidelberg University, Mannheim, Germany*

*Address of correspondence:
Department Medical Physics in Radiology
German Cancer Research Center (DKFZ)
Im Neuenheimer Feld 280
D-69120 Heidelberg
Phone: +49 6221 42 2410
Fax: +49 6221 42 2585
Email: t.kuder@dkfz.de

**Abstract**

This paper presents a novel approach on solving the phase problem in nuclear magnetic resonance (NMR) diffusion pore imaging, a method, which allows imaging the shape of arbitrary closed pores filled with an NMR-detectable medium for investigation of the microstructure of biological tissue and porous materials. Classical q-space imaging composed of two short diffusion-encoding gradient pulses yields, analogously to diffraction experiments, the modulus squared of the Fourier transform of the pore image which entails an inversion problem: An unambiguous reconstruction of the pore image requires both magnitude and phase. Here, the phase information is recovered from the Fourier modulus by applying a phase retrieval algorithm. This allows omitting experimentally challenging phase measurements using specialized temporal gradient profiles. A combination of the hybrid input-output algorithm and the error reduction algorithm was used with dynamically adapting support (shrinkwrap extension). No a priori knowledge on the pore shape was fed to the algorithm except for a finite pore extent. The phase retrieval approach proved successful for simulated data with and without noise and was validated in phantom experiments with well-defined pores using hyperpolarized xenon gas.

**I. Introduction**

Porous media include a wide range of systems such as biological tissue, pore spaces of reservoir rocks filled with fluids [1, 2], or chemical catalysts [3]. Inferring information on the pore structure at the sub-micrometer scale is highly desirable in many areas including life sciences, petrophysics or chemical engineering, but it is generally not easily possible. In nuclear magnetic resonance (NMR) imaging, the self-diffusion of water molecules is widely probed to gain information about the microstructure of porous media [4-6], which would remain concealed in magnetic resonance imaging (MRI) exams otherwise.

Most commonly, diffusive water molecule motion, which is restricted by barriers such as cell membranes, is described by the symmetric diffusion tensor, which quantifies molecular mobility along each axis and is thus indirectly linked to tissue microstructure. This enables the computation of quantitative maps of the apparent diffusion coefficient and fractional anisotropy [7, 8], a parameter characterizing the degree of directionality of the diffusive motion. These diffusion metrics allow differentiation between acute stroke lesions and subacute infarcted areas [9, 10], identification of malignant tumors [11-14] or reconstruction of the 3-dimensional architecture of brain white matter fiber bundles [15-18].

They provide, however, little information about the actual microstructure restricting the diffusion process although such information would be highly valuable in many circumstances. It is known that such information can be obtained by means of "$q$-space imaging", which makes use of the application of two short bipolar gradient pulses. It was used in 1991 by Callaghan et al. to probe the restricted diffusion in closely packed polymer spheres, where they observed an echo attenuation highly reminiscent of diffraction patterns [19]. The signal showed a peak at the wave-vector value corresponding to the inverse of the pore spacing. The replication of the sample's geometry in the signal evolution induced by restricted diffusion lead to the cognition that NMR $q$-space imaging is akin to scattering experiments. Later, diffusion-diffraction patterns outside of phantoms were reported for samples such as erythrocytes [20-22]. For continuous media, the relation between diffusion NMR and the scattering formalism has been outlined in [23, 24].

One limitation of $q$-space imaging is that these diffusion diffraction experiments allow only measuring the modulus squared of the Fourier transform $|\tilde{\rho}(\boldsymbol{q})|^2$ of the unknown pore space function $\rho(\boldsymbol{x})$, which is used to describe arbitrarily shaped pores filled with an NMR-detectable diffusing medium. Hence, the diffraction pattern cannot be directly inverted via Fourier transform to yield the precise pore shape in the rotationally asymmetrical case where the pore's Fourier transform may be negative and/or complex.

The phase problem was overcome by modifying the temporal gradient profile: Laun et al. replaced one of the narrow gradient pulses with a very long pulse of equal area and could thus preserve the phase information [25, 26]. With the greater information content of the thus determinable $\tilde{\rho}(\boldsymbol{q})$, the direct reconstruction of arbitrary pore space functions was enabled so that the average pore shape in an imaging volume element could be measured [27-31]. This technique, known as diffusion pore imaging, potentially enables *in-vivo* estimations of histology-like parameters such as cell-size distributions. However, the need for application of a long gradient pulse prevents the use of many NMR sequences, such as those based on stimulated echoes [32]. Therefore, the alternative solution to the phase problem offered by a special case of double diffusion encoded (DDE) [33] measurements can be beneficial: When using three short gradient pulses, imaginary signal parts can occur [34]. In a subsequent reconstruction of the Fourier transform of the pore space function, the phase information can be extracted from the DDE signal [35, 36]. The magnitude information is either obtainable from the DDE signal or by an additional $q$-space measurement [36, 37]. The feasibility of this approach was demonstrated in well-controlled phantom experiments [38], but robust measurement of the phase of the diffusion-weighted signal is in general a challenging task, particularly *in vivo* [39].

In many fields of physics, it has been shown that lack of phase information can be overcome by means of phase retrieval algorithms. For example, phase retrieval algorithms were used for wavefront sensing for radio antennas [40] and for turbulence-aberrated optics [41], where it was applied to evaluate and correct the aberrations in the Hubble Space telescope [42] and used to align mirror segments of the future James Webb Space telescope [43]. Through solving the phase problem in x-ray crystallography [44], the double helix structure of DNA was uncovered [45]. In lensless imaging, phase retrieval algorithms are used as a substitute of the lens to recombine the scattered x-ray light offering aberration-free diffraction-limited images [46, 47]. 3D images can be constructed tomographically [48, 49] and application to biological samples is feasible [50-52]; for more application areas see [53].

In this work, we propose to apply phase retrieval algorithms to $q$-space imaging data to reconstruct pore shapes. The phase information is retrieved by an iterative process from the available magnitude information using additional conditions, in particular assuming that the imaged pores are of finite size while employing a dynamic support estimation [54]. Validation using diffusion simulations and measurements in well-defined geometries using hyperpolarized xenon-129 is presented.

## II. Theory

### A. The pore space function

A pore is understood to be a finite volume $V$ that has closed boundary and whose interior is defined by the support $\Omega$. Its volume shall be filled with an NMR-detectable diffusing medium. The pore shape is described by the pore space function

$$\rho(x) = \begin{cases} 1/V, & \text{if } x \in \Omega \\ 0, & \text{if } x \notin \Omega. \end{cases} \quad (1)$$

In diffusion pore imaging, one is interested in determining $\rho(x)$ by measurement of its Fourier transform $\tilde{\rho}(q)$, where $q$ represents a vector in $q$ space and $|q|$ is referred to as the $q$ value.

### B. q-Space imaging and the phase problem

In NMR diffusion experiments, the signal attenuation is given by the ensemble average $\langle \exp(i\psi) \rangle$ over all possible random walk trajectories, where $\psi = -\gamma \int_0^T G(t) \cdot x(t) dt$ is the acquired spin phase by a random walker traversing on the trajectory $x(t)$ during $0 < t < T$ in the presence of the temporal magnetic field gradient profile $G(t)$, and $\gamma$ is the gyromagnetic ratio [55]. For $q$-space imaging, two short bipolar gradient pulses of duration $\delta$ are applied with gradient vectors $-G$ at $t = 0$ and $G$ at $t = T - \delta$. For a spin echo version of the gradient profile see Fig. 1(a). With this definition, immobile particles accumulate no net phase whereas moving particles cause a signal attenuation given by

$$S_{11}(q) = \left\langle \exp\left[-iq \cdot \left(-\frac{1}{\delta}\int_0^\delta x(t)dt + \frac{1}{\delta}\int_{T-\delta}^T x(t)dt\right)\right] \right\rangle \quad (2)$$

with $q = \gamma G \delta$. In the limit of narrow gradient pulses ($\delta \to 0$), the time integrals equal the particles starting position $x_1 = x(0)$ and final position $x_2 = x(T)$, and in the limit of long diffusion time ($T \to \infty$), each exponential can be evaluated separately because the correlation between $x_1$ and $x_2$ is lost [19, 56, 57]:

$$S_{11}(q) = \langle e^{iq \cdot x_1} \rangle \langle e^{-iq \cdot x_2} \rangle = \int_{\text{Pore}} \rho(x_1) e^{iq \cdot x_1} dx_1 \int_{\text{Pore}} \rho(x_2) e^{-iq \cdot x_2} dx_2$$

$$= \tilde{\rho}^*(q)\tilde{\rho}(q) = |\tilde{\rho}(q)|^2, \quad (3)$$

where the asterisk marks the complex conjugate. In Eq. (3), the phase information on the form factor $\tilde{\rho}(q)$ is absent so that $q$-space imaging only allows measuring the magnitude spectrum. Hence, an unambiguous determination of $\rho(x)$ from $q$-space measurements via Fourier transform is impossible, which is known as the phase problem.

*C. Phase extraction from phase-containing measurement: Double diffusion encoding (DDE)*

The phase information can be recorded using two diffusion encodings with antiparallel wave vectors [34] applied without temporal separation so that the second and third gradient pulse are superimposed [Fig. 1(b)]. For this special form of double diffusion encoding [33], the signal in the short-gradient-pulse approximation is given by [34, 35]

$$S_{121}(\boldsymbol{q}) = \tilde{\rho}^*(\boldsymbol{q}/2)^2 \tilde{\rho}(\boldsymbol{q}). \tag{4}$$

The desired phase of $\tilde{\rho}(\boldsymbol{q})$ is not directly accessible but can be disentangled from Eq. (4) with an iterative phase estimation approach [36, 38]. The phase can either be combined with the magnitude information from $q$-space imaging or the full complex signal in $q$ space can be obtained solely from the double diffusion encoded measurement when the magnitude is as well estimated from Eq. (4).

*D. Phase retrieval*

Using phase retrieval, a function, here the pore space function $\rho(x)$, is recovered from the magnitude measurement of its Fourier transform, $|\tilde{\rho}(\boldsymbol{q})|$, or, equivalently, $\varphi(\boldsymbol{q}) = \arg[\tilde{\rho}(\boldsymbol{q})]$ is retrieved from $|\tilde{\rho}(\boldsymbol{q})|$. Without further information on $\rho(x)$, the phase retrieval problem is ill-posed. A better posed problem can be obtained by imposing additional properties on the image space candidate solution such as compact support constraint and that $\rho(x)$ is real. With such additional information, $\rho(x)$ can be uniquely determined up to trivial ambiguities [58, 59]. Even in settings for which uniqueness is guaranteed, no general phase retrieval method exists to recover the unknown phase. The most popular class of phase retrieval methods are alternating projection algorithms which are based on the pioneering work by Gerchberg and Saxton [60], who presented the first iterative method, capable of phasing diffracted intensities measured in the object and Fourier domain. This approach was extended by Fienup who swapped the magnitude information in the object's domain against feedback and compact support constraints [61]. He proposed different versions with differing constraints imposed on the current image estimate: two of these are the error reduction algorithm (ER) [62] and the hybrid input-output algorithm (HIO) [62].

The basic outline of the ER algorithm is as follows: First, an initial random guess of the pore image $\rho_1(x)$ is made, and then repeated Fourier transforms between image and $q$ space are performed after the known information on the pore image to be reconstructed are imposed on the current estimate in each domain. In Fourier space, consistency with the measured magnitude $|\tilde{\rho}(q)|$ is imposed. In the beginning, a random phase is assigned to the known magnitude. In image space, the pore is constrained to have a finite support $\Omega$, i.e., there is an area outside the pore which is identically zero. Assuming that the pore is a finite-size object is an essential step in solving the phase problem. The algorithm can be formulated mathematically in four simple steps, which are repeated iteratively with $k$ as the iteration index:

Step 1: $\quad \tilde{\rho}_k(q) = |\tilde{\rho}_k(q)| \exp[i\varphi_k(q)] = \text{FT}\{\rho_k(x)\}$ (5)

Step 2: $\quad \tilde{\rho}'_k(q) = |\tilde{\rho}(q)| \exp[i\varphi_k(q)]$ (6)

Step 3: $\quad \rho'_k(x) = |\rho'_k(x)| \exp[i\vartheta'_k(x)] = \text{FT}^{-1}\{\tilde{\rho}'_k(q)\}$ (7)

Step 4 (ER): $\quad \rho_{k+1}(x) = \begin{cases} \rho'_k(x), & \text{if } x \in \Omega \\ 0, & \text{if } x \notin \Omega \end{cases}$ (8)

For better readability, a block diagram of the algorithm is shown in Fig. 2 with notation adapted to NMR diffusion pore imaging [63]. In the beginning, the current estimate $\rho'_k(x)$ is in general incorrect. However, when the calculated Fourier magnitude $\tilde{\rho}_k(q)$ is substituted with the measured magnitude $|\tilde{\rho}(q)|$ in step 2 while the calculated phase is kept the same, the inverse Fourier transform (step 3) yields a refined estimate of the pore image $\rho'_k(x)$. After correcting the estimate by imposing a boundary and setting all elements outside the boundary to zero (step 4), transforming the image again into Fourier space (step 1) provides a refined estimate of the phase $\varphi_k(q)$, which is again combined with the measured magnitude for the next iteration. As the iterations proceed, the pore image estimate will converge to the correct image that satisfies the constraints in both $x$ and $q$ space.

It is known that the error reduction algorithm is prone to stagnation in local minima of the error between measurement $|\tilde{\rho}(q)|$ and estimate $|\tilde{\rho}_k(q)|$ [59]. To avoid stagnation in a local minimum, Fienup proposed the HIO version [61, 63], for which the first three steps are identical to the ER version but it differs in the fourth: The outer region, where $|\rho'_k(x)|$ should converge to zero, is suppressed by taking the result of the previous iteration $\rho_k(x)$ into account and subtracting the new estimate from the current iteration times a feedback coefficient $0 \leq \beta \leq 1$. Typical values for $\beta$ range between 0.5 and 0.9 [53, 63]. Equation (8) is replaced by

$$\text{Step 4 (HIO):} \quad \rho_{k+1}(x) = \begin{cases} \rho'_k(x) & , \text{ if } x \in \Omega \\ \rho_k(x) - \beta \rho'_k(x), & \text{ if } x \notin \Omega. \end{cases} \tag{9}$$

By means of the negative feedback component, the starting point for the next iteration is pushed into the desired direction and the algorithm is able to escape from local minima of the error metric, i.e., the errors can increase temporarily and there is no proof of convergence.

To help the algorithm to converge, other prior information on the sought-after pore space function can be incorporated, for example imposing a non-negativity (NN) constraint on $\rho'_k(x)$ in addition to requiring a compact support:

$$\text{Step 4 (ER+NN):} \quad \rho_{k+1}(x) = \begin{cases} \rho'_k(x), & \text{if } (x \in \Omega) \wedge [\rho'_k(x) \geq 0] \\ 0 & , \text{ if } (x \notin \Omega) \vee [\rho'_k(x) < 0], \end{cases} \tag{10}$$

$$\text{Step 4 (HIO+NN):} \quad \rho_{k+1}(x) = \begin{cases} \rho'_k(x) & , \text{ if } (x \in \Omega) \wedge [\rho'_k(x) \geq 0] \\ \rho_k(x) - \beta \rho'_k(x), & \text{ if } (x \notin \Omega) \vee [\rho'_k(x) < 0]. \end{cases} \tag{11}$$

Negativity refers to the real part of $\rho'_k(x)$. The imaginary part is ignored.

When applying phase retrieval to $q$-space data to perform diffusion pore imaging, the pore boundary is the main unknown information to be determined and at the same time a compact support is essential to solve the phase problem. Therefore, $\Omega$ needs to be deduced from information that is available in the measurement. Back-transforming the $q$-space measurement, Eq. (3), results in

$$\text{FT}^{-1}\{\tilde{\rho}^*(q)\tilde{\rho}(q)\} = \text{FT}^{-1}\{\tilde{\rho}^*(q)\} * \text{FT}^{-1}\{\tilde{\rho}(q)\} = \rho(-x) * \rho(x) =: A(x), \tag{12}$$

where the convolution $\rho(-x) * \rho(x)$ is known as the pore's autocorrelation function $A(x)$, which is non-zero over a range of twice the pore's extent for each of both dimensions. By thresholding $A(x)$ appropriately, a first loose support of the pore is found, which is large enough to comprise the pore but does generally not allow for an exact determination of the support. One solution to this problem is the shrinkwrap extension that has been developed to find the support dynamically [54, 64]: While the algorithm is progressing, the support will be tightened by thresholding a blurred version of $\rho'_k(x)$ until it wraps around the actual pore shape (Fig. 2). This way, the shape of the pore is determined together with the image of the pore.

The squared Fourier modulus is insensitive to multiplicative constant phase factors $e^{i\alpha}$ for $\alpha \epsilon [0, 2\pi)$, translations of $\rho(x)$ by some $x_0 = (x_0, y_0)$ or to conjugate reversal (complex conjugation plus rotation by 180°, i.e. $\rho(x) \to \rho^*(-x)$), with the latter being referred to as image twinning: $|\text{FT}\{\rho\}(q)|^2 = |\text{FT}\{\psi\}(q)|^2$ if $\psi = e^{i\alpha}\rho^*(-x - x_0, -y - y_0)$. In phase retrieval, these ambiguous solutions are considered to be equivalent.

*E. Pore shape and size distributions*

The total $q$-space signal over $M$ different separated pores with different pore space functions $\rho_n(x)$ and pore volumes $V_n$ is given by [65]

$$S_{11,\text{tot}}(q) = \sum_{n=1}^{M} f_n |\tilde{\rho}_n(q)|^2 \qquad (13)$$

with the volume fraction of the NMR-detectable medium $f_n = V_n / (\sum_{n=1}^{M} V_n)$. The form factor does not contribute linearly to the total signal but instead the product with its complex conjugate, $\tilde{\rho}^*(q)\tilde{\rho}(q)$, enters. Thus recording the $q$-space signal over a volume element with different pores does not return an arithmetic average pore image. In contrast, for the long-narrow gradient scheme [25], where changing one of the short $q$-space gradient pulses to a long pulse with identical first moment preserves the phase information, the form factor appears linearly in the respective signal equation [65] ($\delta \to 0, T \to \infty$):

$$S_{\text{LN,tot}}(q) = \sum_{n=1}^{M} f_n \tilde{\rho}_n(q) \exp[i q \cdot x_{n,\text{cm}}].$$

As a result of the factor $\exp[i q \cdot x_{n,\text{cm}}]$, where $x_{n,\text{cm}}$ is the respective pore's center of mass, all pores in the imaging volume are shifted on top of each other. Under ideal conditions ($\delta \to 0, T \to \infty$), an average pore image is obtained by taking the inverse Fourier transform of $S_{\text{LN,tot}}(\boldsymbol{q})$. For the derivation of $S_{\text{LN}}(\boldsymbol{q})$ analogous to section II B, the reader is referred to [25, 26].

## III. Methods

*A. q-Space, DDE and long-narrow data*

*1. Simulations of uniform samples*

Numerical simulations of the diffusion process within confining triangular pore shapes were performed as in [38]: The $q$-space and double diffusion encoded signals were computed using an eigenvalue decomposition approach (using Eq. 144 of [66], see also Refs. [26, 67-70]). The expense on computational time is very low with this approach but it requires that the Laplacian eigenfunctions and matrices are known analytically for the pore domain, as is the case for the equilateral triangular pore shape [26]. It was ensured that the used number of eigenvalues was sufficient by varying the number of eigenvalues. The edge length of the triangular shape was $L = 3400$ μm. The diffusion time was $T = 270$ ms and the gradient duration was $\delta = 5.46$ ms. For phase retrieval, 25×25 points in $q$ space were sampled with maximum $q$ values $q_{\max}$ of ± 12 mm$^{-1}$ in both vertical and horizontal dimensions corresponding to a nominal resolution of $\Delta x = 0.26$ mm. For the recursive reconstruction, $q$-space and DDE signals were sampled at a spacing of 0.12 mm$^{-1}$ for the vertical and horizontal gradient direction.

For star-shaped pores, Monte Carlo simulations with $1.5 \times 10^6$ random walkers and $6 \times 10^4$ steps per random walk trajectory were used to compute the q-space signal. Parameters used: Pore size as in Fig. 1(d), 27×27 sampling points in $q$ space, $q_{\max} = \pm 13.5$ mm$^{-1}$, $\Delta x = 0.23$ mm, $T = 340$ ms, $\delta = 6.15$ ms.

The effect of a limited signal-to-noise ratio (SNR) was estimated by adding Gaussian noise with a standard deviation equal to 1/150 times the signal at $q = 0$ to the signal $S_{11}(\boldsymbol{q})$.

*2. Simulations of pore distributions*

To assess the performance of the phase retrieval method for imaging volumes containing not identically shaped or sized pores in comparison to the long-narrow approach, simulations were performed for four different settings: In the first setting, the $q$-space imaging and long-narrow signal of a shape distribution containing the equilateral triangle and the star shape were simulated with the following parameters: Triangle and star size as in Fig. 1(c-d), $T = 340$ ms, $\delta = 6.15$ ms, 27×27 points, $q_{max}=\pm 13.5$ mm$^{-1}$, $\Delta x = 0.23$ mm. Further, two homogenous size distributions of 101 triangles with mean size $L_0 = 3400$ μm ($L = 0.75\,L_0 \ldots 1.25\,L_0$ for a narrow distribution, $L = 0.15\,L_0 \ldots 1.85\,L_0$ for a broad distribution) were simulated as well as one orientation distribution of identically sized triangles ($L = 3400$ μm), which were rotated in steps of 1° from -20° to 20°. Parameters for the three settings: $T = 270$ ms, $\delta = 5.46$ ms, 27×27 points, $q_{max}=\pm 12$ mm$^{-1}$, $\Delta x = 0.26$ mm. Additionally, for all distributions the volume fraction-weighted average pore space function under idealized conditions ($\delta \to 0, T \to \infty$) was computed analytically for the triangular domains or from the Monte Carlo simulation for the star shape.

## 3. Experiments

All experiments were conducted on a clinical MR scanner of 1.5 T static magnetic field (Magnetom Symphony, A Tim System, Siemens Healthcare, Erlangen, Germany) with a maximal employed gradient amplitude of 29.5 mT/m. $q$-Space diffusion measurements require the diffusion process to reach the long-time limit during the diffusion time, while the typical diffusion distance during gradient application has to be small compared to the pore size imposing high requirements on the gradient amplitude. As previously described, both requirements can be met on a clinical MR scanner by using phantoms containing pores on the millimeter scale filled with a hyperpolarized xenon-129 gas mixture [27, 38]. The xenon diffusion coefficient of the mixture containing Xe (0.95 Vol %), N$_2$ (8.75 Vol %) and He (rest) was estimated to $D_0 = (37\,000 \pm 2\,000)$ μm²/ms [27], which is almost one order of magnitude higher than for pure xenon gas [71]. Hyperpolarized gas was generated using spin exchange optical pumping and was transferred to the phantom in the MR scanner as detailed in [27, 38].

Diffusion pore imaging phantoms with two different pore shapes were studied: The first type consists of acrylic glass plates with cutout grooves of equilateral triangular shapes and was built by the in-house workshop. The plates were stacked on top of each other to form an array of 170 pores [Fig. 1(c)]. The second phantom was 3D-printed as one block with the PolyJet technology (Objet30 Pro, VeroClear as printing material, Stratasys, Ltd., Eden Prairie, MN, USA) and contained 135 tubes with star shaped cross-section as illustrated in Fig. 1(d). These phantoms were positioned in the isocenter of the magnet in an in-house built xenon coil, and a constant gas flow of 140 ml/min through the pores was directed parallel to the main magnetic field.

The radiofrequency pulses were applied to select one thick slice of 45 mm with the plane normal vector parallel to the gas flow covering the complete length of the pore tubes. This way, the diffusion encoding gradients were applied in the transversal plane of the scanner orthogonal to the flow direction, i.e. in the plane of the triangular and star shaped cross-sections. $q$ space was sampled on a Cartesian grid using the $q$-space gradient profile depicted in Fig. 1(a) by varying the gradient amplitudes while keeping the gradient duration $\delta$ constant. For triangular pores the gradient duration was $\delta = 5.46$ ms and the diffusion time was $T = 270$ ms. For the maximum $q$ value ± 12 mm$^{-1}$ with $\Delta q$ = 1 mm$^{-1}$ were used corresponding to a nominal resolution of $\Delta x$ = 0.26 mm and 25×25 pixels. For the star-shaped pores $T = 340$ ms and $\delta = 6.15$ ms were used with $q_{\max}$ = ± 13.5 mm$^{-1}$, $\Delta q$ = 1.039 mm$^{-1}$, $\Delta x$ = 0.23 mm and 27×27 pixels. For the triangular pores, additional double diffusion encoded measurements were conducted for the vertical and horizontal gradient direction using the profile depicted in Fig. 1(b) with $\Delta q$ = 0.48 mm$^{-1}$. Both gradient profiles were implemented as spin echo sequences by inserting 180° refocusing pulses with durations of 2.56 ms, which were surrounded by spoiler gradients in slice selection direction. The durations $\delta$ of the trapezoidal gradient pulses are given as flat top time plus the ramp up time of 0.30 ms. The repetition time, i.e. the time between two consecutive 90° excitation pulses, was set to 18 s to restore the polarization in the phantom sufficiently via gas exchange before a new point in $q$ space was acquired. Fluctuations in the polarization level were accounted for using pre-readouts: The recorded spin echo was averaged and normalized to an additional signal acquisition directly after the 90° excitation pulses. To obtain the diffusion-induced signal attenuation, a normalization to $S(\boldsymbol{q} = 0)$ was conducted afterwards. For phase retrieval, the absolute value of the $q$-space signal was used.

For the star-shaped domain, due to limited SNR at the higher $q$ values, the $q$-space signal was measured three times and was then averaged before applying the phase retrieval algorithm.

*B. Phase retrieval algorithm: Initialization and used parameters*

In this work, the algorithm consists of several cycles of iterations, where one cycle consists of many hundreds of iterations of the HIO algorithm followed by a few iterations of the ER algorithm while incorporating a dynamically adapting support using the shrinkwrap extension. The first estimate of the support mask was obtained by thresholding the autocorrelation function $A(\boldsymbol{x})$ [Eq. (12)] at 5 % of its maximum so that the up-right pore and its twin image fit inside. Every 10 iterations of the algorithm (HIO or ER), Ω was improved by convolving the magnitude of the current estimate $\rho'_k(\boldsymbol{x})$ with a Gaussian kernel and thresholding this blurred version at 20 % of its maximum. The width (standard deviation) of the Gaussian, initially set to $\sigma = 2.5$ pixels, was reduced at each update of the support by 2 % so that the support shrank from update to update until it enclosed the pore shape tightly. At a minimum of 0.5 pixels, $\sigma$ was kept constant.

To begin, an array of random numbers in the interval (0,1) was generated to serve as an initial unbiased estimate of the pore image $\rho_1(\boldsymbol{x})$, which was normalized so that its Fourier transform had the same integral as the measured Fourier magnitude data, i.e., $\sum_i \sum_j |\tilde{\rho}_{1,ij}| = \sum_i \sum_j |\tilde{\rho}_{ij}|$ where $\tilde{\rho}_{1,ij}$ and $\tilde{\rho}_{i,j}$ denote the respective elements of the image matrix. This normalization was also performed on $\tilde{\rho}_k(\boldsymbol{q})$ at every iteration. Since $\rho(\boldsymbol{x})$ is real per definition, a global phase was additionally multiplied to every pixel of $\tilde{\rho}_k(\boldsymbol{q})$ to zero the phase in the central pixel at every iteration. 2000 iterations of the HIO algorithm with non-negativity constraint [Eq. (11)] with $\beta = 0.9$ were performed followed by 300 iterations of the ER algorithm with non-negativity constraint [Eq. (10)]. The shrinkwrap refinements started with the first iteration of the HIO algorithm. After the last HIO iteration, the newest version of $\Omega$ and the current $\sigma$ were passed on to the ER algorithm. The sequence of random start generation and HIO iterations followed by ER iterations is referred to as one cycle of the phase retrieval algorithm.

For each tested cycle, starting each from a random array, the reconstructed image and the shape of its shrinkwrap-support slightly varied. Therefore 100 cycles were carried out and averaged: To remove translational ambiguity and image twinning, each of the 100 pore images was moved to the image center. As reference, the image featuring the highest asymmetry was chosen, i.e. the one with the largest difference to its rotated version. The other pore images were matched to this reference by rotation, if rotation decreased the deviation from the reference. In this work, the average pore image together with the phase of the inverse Fourier transform of the average pore image are referred to as the phase retrieval result.

### *C. Processing of the DDE signal*

To assess and compare the phase retrieval result, the phase was calculated for the vertical and horizontal gradient direction in $q$ space from simulated and measured double diffusion encoded signals [Eq. (4)] through a recursive reconstruction approach as described in [36, 38]. Afterwards, an inverse Fourier transform was applied to both directions returning the projection of the pores onto the respective gradient direction.

Simulations, signal processing, the phase retrieval algorithm and the iterative phase reconstruction from the DDE signal were implemented in Matlab (MathWorks, Natick, MA, USA).

## IV. Results

Figure 3 shows the phase retrieval result for the signals obtained in diffusion simulations (III.A.1) using a realistic gradient timing for a clinical scanner ($\delta$ =5.46 ms, $T$ =270 ms). It demonstrates that the correct pore shape was found by the phase retrieval algorithm, as well as pixel, that are crossed by the pore boundary and have low intensity due to partial volume effects [Fig. 3(a)]. Looking at individual gradient directions, the algorithm was successful to the same degree as the recursive phase reconstruction from the simulated double diffusion encoded (DDE) signal: Figures 3(d,e) show the profiles of $\tilde{\rho}'_k(\mathbf{q})$ for the vertical (d) and horizontal (e) gradient direction, indicated by dots, and compare them to profiles where the phase information was disentangled from the simulated DDE signals, indicated by lines. In Fig. 3(f,g), the projection of $\text{Re}[\rho'_k(x)]$ onto the respective gradient direction is shown (dots) and compared to the one-dimensional inverse Fourier transforms of the $\tilde{\rho}(q)$-profiles from DDE simulations (lines). For the horizontal gradient direction, the retrieved phase is identical to zero at each $q$ value resulting in a purely real $\tilde{\rho}_k(q)$ as expected for the mirror-symmetric $x$-space profile. For the vertical gradient direction, the phase is not zero leading to a complex $\tilde{\rho}_k(q)$. The algorithm correctly reproduces $\tilde{\rho}_k(\mathbf{q}) = \tilde{\rho}_k^*(-\mathbf{q})$ resulting in sign changes of the phase in [Fig. 3(c)] when replacing $\mathbf{q}$ by $-\mathbf{q}$, which is also confirmed in Fig. 3(d). Both $x$-space profiles clearly show the projection of the pore onto the two directions. The reconstructed pore image shows a signal increase near the boundary [Fig. 3(a)]: Since the gradient pulse duration is not negligible, the center of mass of the trajectory of a particle diffusing near the boundary during the time interval at which a gradient is applied, will not lie directly next to the boundary but will be shifted toward the center of the pore causing a shift of the diffraction maxima to higher $q$ values and the pore appears smaller than its real size with increased boundary signal. This edge enhancement effect [26, 38, 72] can also be seen clearly in (f).

For the same $q$-space data as in Fig. 3, Fig. 4 shows a sequence of images produced by the phase retrieval algorithm as the image reconstruction progresses, starting with the magnitude-only information in $q$ space, the initial random image in $x$ space and the support from the autocorrelation function, which is shaped like a hexagon. For an increasing number of iterations the random pore image estimate transforms into the triangular shape: The phase profile improves steadily and the support shrinks and wraps around the pore shape tightly. Here, both a translation and image twinning occurred so that the pore appears upside down.

Figure 5 demonstrates that the phase retrieval algorithm is very robust to noise when we set out to find the pore shape in a uniform sample (SNR = 150). However, details such as the enhanced edges as for the simulation without noise or the fine triangle tips at the bottom [Fig. 3(a)] cannot be recovered unambiguously at this noise level.

Moving on to experimental data (III.A.2), the diffraction pattern [Fig. 6(b)] shows considerable noise at the higher $q$ values and in areas where the signal decreases faster, i.e., where the $x$-space projections onto the gradient directions are mirror-symmetric. At higher $q$ values, these are also the areas, where the retrieved phase becomes inaccurate and the conjugate symmetry gets lost. However, for the vertical and horizontal gradient direction, the profiles in $q$ and $x$ space (dots) agree very well with profiles extracted from simulated (lines) DDE signals [Fig. 6(d-g)]. For the measured DDE signals (crosses in Fig. 6), twice the number of $q$ values was acquired since the recursive phase estimation in section II.C takes the phase of the form factor at half the $q$ value into account, and therefore involves interpolation between measurement points [36, 38]. Here, stronger deviations at high $q$ values from the simulations and the phase retrieval result are observed: $S_{121}(q)$ drops relatively fast to the noise level. Therefore, the recursive phase estimation using the already noisy signal at smaller $q$ values fails at high $q$ values (for some directions) [38]. The phase retrieval algorithm found the correct pore shape and also the edge enhancement effect is not concealed by noise [Fig. 6(a)].

Figure 7 illustrates that phase retrieval from $q$-space imaging data is feasible for a variety of pore shapes as shown here exemplary for star-shaped pores. For the simulated $q$-space signal, the corners of the $q$-space magnitude image have very low signal values such that, even for the simulation, the corners of the phase image cannot be retrieved reliably [Fig.

7(c)] due to Monte Carlo noise. Nonetheless, the star shape was reconstructed successfully [Fig. 7(a)]. The tips of the angles show an increased signal due to the edge enhancement effect. Although the signal was measured thrice to improve SNR, only the first ring of diffraction peaks around the central maximum could be clearly observed, while the second ring is lost in noise, so that the phase could be retrieved successfully only at small $q$ values [Fig. 7(f)]. This resulted in a blurred version of the star with somewhat less distinct angle tips and slightly reduced size in Fig. 7(d) compared to (a), but a distinct star shape could still be formed.

The question of applicability of the phase retrieval method to pore distributions is addressed in comparison to the long-narrow approach in Fig. 8. For the case of two different pore shapes contained in the imaging volume, here triangle and star, the tips of triangle and star are more pronounced and sharper for the phase retrieval method compared to the long-narrow approach, so that the two individual shapes can be better recognized in the phase retrieval result [Fig. 8(a)]. The true distribution is depicted in the left column for comparison. While the intensity distribution for the long-narrow approach is homogenous in the center where both shapes overlap, the phase retrieval result shows a low intensity spot at the central bottom of the triangle. For both size distributions, phase retrieval returns an image of a single pore of mean size [Fig. 8(b-c)]. But both of the two compared methods do not allow on their own to distinguish, whether the imaging volume contained a distribution of different sizes or only triangles of one size in-between the smallest and largest occurring size. For the broader distribution, artifacts occur at the triangles tips for the phase retrieval method and the long-narrow approach experiences heavy blurring. In (d) the phase retrieval method results in a better representation of the orientation distribution.

## V. Discussion

In this work, NMR diffusion pore imaging was extended by a new approach: The phase retrieval methodology proved successful in solving the phase problem in diffusion pore imaging for simulated and experimental data. This new approach differs from previous approaches in a fundamental way. Whereas previous approaches employed explicit phase measurements [27, 28, 31, 38], the new approach uses classical single diffusion encoding which by itself only provides magnitude information. The novelty lies in recovering the full complex signal by applying a phase retrieval algorithm, which is well known from analogous phase problems in various fields of physics and eliminates the need for experimentally challenging phase measurements. Recovery of the missing phase information in the measured signal $S_{11}(\boldsymbol{q}) = |\tilde{\rho}(\boldsymbol{q})|^2$ was only made possible by including the knowledge of $\rho(\boldsymbol{x})$ being zero outside a finite domain.

However, there is no general algorithm guaranteeing recovery of the true phase of the form factor $\tilde{\rho}(\boldsymbol{q})$. In this work, a combination of the HIO algorithm and the ER algorithm was used and complemented with the shrinkwrap extension. One run of this combination with specific random start image $\rho_1(\boldsymbol{x})$ is referred to as one cycle. Using only one cycle proved to be robust for triangular domains without noise. An important issue is the sensitivity of the phase retrieval algorithm to inaccurate information on the pore space function to be reconstructed, i.e., when the modulus of the Fourier transform is not exactly known due to measurement noise. If the Fourier modulus is affected by noise, each reconstructed pore image, each seeded with a different random start, will vary slightly since there is no true solution that exactly fulfills Fourier and image domain constraints. In the presence of noise, not each tested cycle of the algorithm matched the true pore image perfectly, but generally the triangular and the star shape were clearly recognizable. Averaging the retrieved pore images from 100 independent cycles led to very good results: Comparing simulations, noise-free and affected by noise, and experimental data, we found the phase retrieval algorithm to be robust to noise and capable of revealing pore shapes at lower SNR values than occurring in our experimental data.

It was noticed that convergence to the correct solution can be accelerated, if the additional non-negativity constraint was used. The support plays a key role in finding the true solution: The initial estimate of the support $\Omega$ needs to be large enough so that $\rho(\boldsymbol{x})$ vanishes outside the pore. Thresholding the autocorrelation function provides a rough guess to determine an upper boundary of the pore size. Using the shrinkwrap extension, the support is tightened around the pore and the initial symmetry of $\Omega$ inherited from the autocorrelation gets broken which makes convergence more likely by ruling out a twin image. If the support region truncates the pore, i.e. does not contain all nonzero pixels, it might cause the algorithm to stagnate; however, overshrinking can usually be corrected by the shrinkwrap algorithm. A slightly larger support offers more freedom for the algorithm to reconstruct boundary pixels that show partial volume effects. The support mask does not necessarily need to trace the exact boundary of the pore, but, in practice, the support should be reasonably tight, because otherwise the pore image estimate might change rapidly between solutions that differ in regard of trivial ambiguities, i.e., translation and image twinning. A tight support also helps to reconstruct intensity variations inside the pore which can occur due to e.g. an inhomogeneous magnetization or media density, or as seen here, due to finite gradients resulting in enhanced edges and blurring. The shrinkwrap mechanism is quite robust, so that, over the course of the iterations, the support might wander outside of the initial autocorrelation support and is even often able to converge if the pore is translated in a way that it crosses the edges of the image array and the support gets fragmented into two or more parts.

Concerning practical implementation of this $q$-space imaging-based method, some limitations have to be noted: When investigating an imaging volume containing pore size or shape distributions, relying on two short gradient pulses only implies that instead of an arithmetical average pore image the average of the products of $\tilde{\rho}(\boldsymbol{q})$ with $\tilde{\rho}^*(\boldsymbol{q})$ is recorded [65, 73] prohibiting a direct inversion to obtain the actual pore distribution. If the pores are not relatively monodisperse in size and have random orientation, the diffraction pattern in the $q$-space signal will vanish making the microstructure information more difficult to access. The relevance of this problem depends strongly on the investigated geometries and distributions. Short gradients cause pore shape specific artifacts such as the underestimated area of the phase retrieval result for the sample containing two different shapes. Not having to consider each case individually, as given for the long-narrow approach, is largely beneficial in practical applications. Nonetheless, phase retrieval yielded a better representation of the shape distribution as well as for the orientation distribution in comparison to the long-narrow approach. The problem of measuring the average of $\tilde{\rho}^*(\boldsymbol{q})\,\tilde{\rho}(\boldsymbol{q})$ was much more relevant for broad size distributions where an inversion to the true distribution seems impossible. It should be noted that also the long-narrow approach does not result in the true average for finite gradients.

Further, the phase retrieval algorithm bases on the assumption that pores are of finite size with a closed boundary and surrounded by a region with no signal contribution. However, in biological tissue, the typical setup is quite different, which makes it a challenge to retrieve a pore image even if the limitation of size and shape heterogeneity is neglected: Cell membranes are very thin and permeable and treating them as solid casings does not seem feasible. In addition, extracellular signal compartments will have to be suppressed which may be performed using filter gradients and high gradient amplitudes are mandatory to perform pore imaging at the micrometer scale [26].

Comparison of the phase retrieval approach to the short-gradient method using recursive phase reconstruction from DDE signals shows a decreased sensitivity to noise for the phase retrieval approach. Moreover, using phase retrieval, the measurement time can be reduced extensively: The DDE pore imaging approach with the lowest sensitivity to noise [38] needs at least twice the sampling points since both $q$-space and DDE measurements are required. Additionally, for the DDE approach, a higher sampling density of the $q$ values is necessary to achieve a stable reconstruction. Particularly in contrast to the long-narrow gradient scheme, which records the full Fourier signal [25, 26], using the standard bipolar gradient shape is advantageous because the application is especially easy and stable from a technical point of view.

Eliminating the need for experimentally challenging phase measurements could in principle be advantageous for *in-vivo* imaging of tissue microstructure, where cardiac driven tissue pulsation and patient movement complicate phase measurements, but have no bearing on $q$-space imaging. This may improve the applicability of diffusion pore imaging in imaging porous structures with a higher SNR compared to conventional MRI.

In conclusion, it was demonstrated in simulations and phantom experiments that diffusion pore imaging is possible using $q$-space imaging data eliminating the need for additional phase measurements using specialized temporal gradient profiles.

## Acknowledgment

Financial support by the DFG (grant no. KU 3362/1-1 and LA 2804/6-1) is gratefully acknowledged.

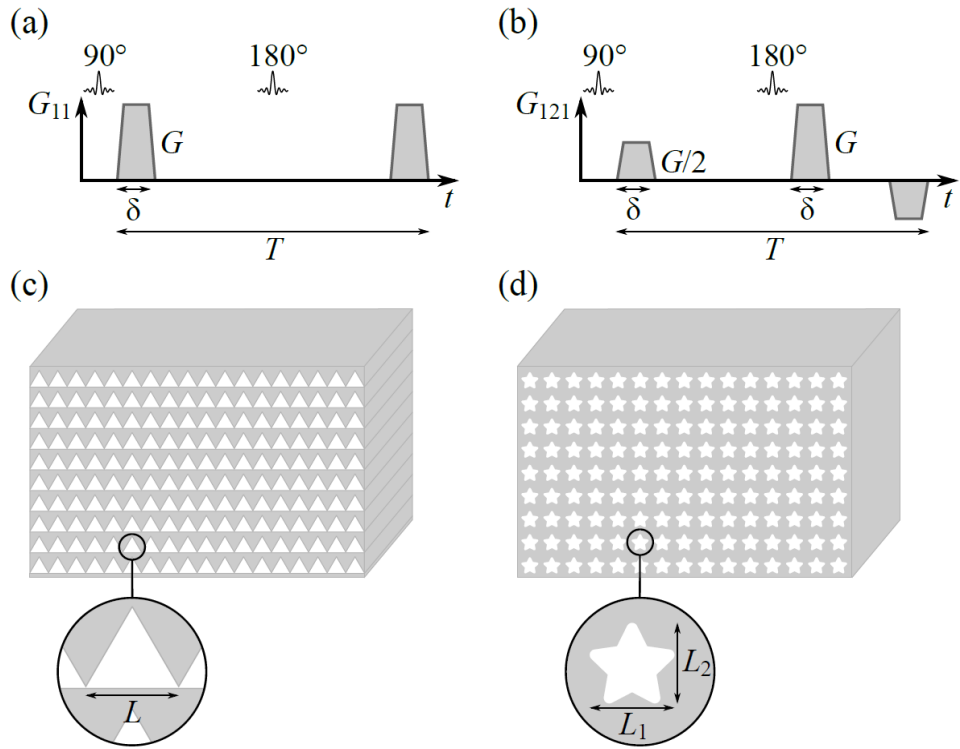

**FIG. 1. (One column)**

Schematic depiction of the diffusion-encoding gradient profiles and pore imaging phantoms. (a) Effective $q$-space imaging gradient profile $G_{11}(t)$. (b) Double diffusion encoding gradient profile $G_{121}(t)$. Gradient durations $\delta$ are composed of the gradient ramp up time plus flat top time. Pore imaging phantoms form arrays of pores with (c) equilateral triangular cross section with edge length $L = 3400$ μm or (d) star shapes with $L_1 = 3100$ μm, $L_2 = 3000$ μm.

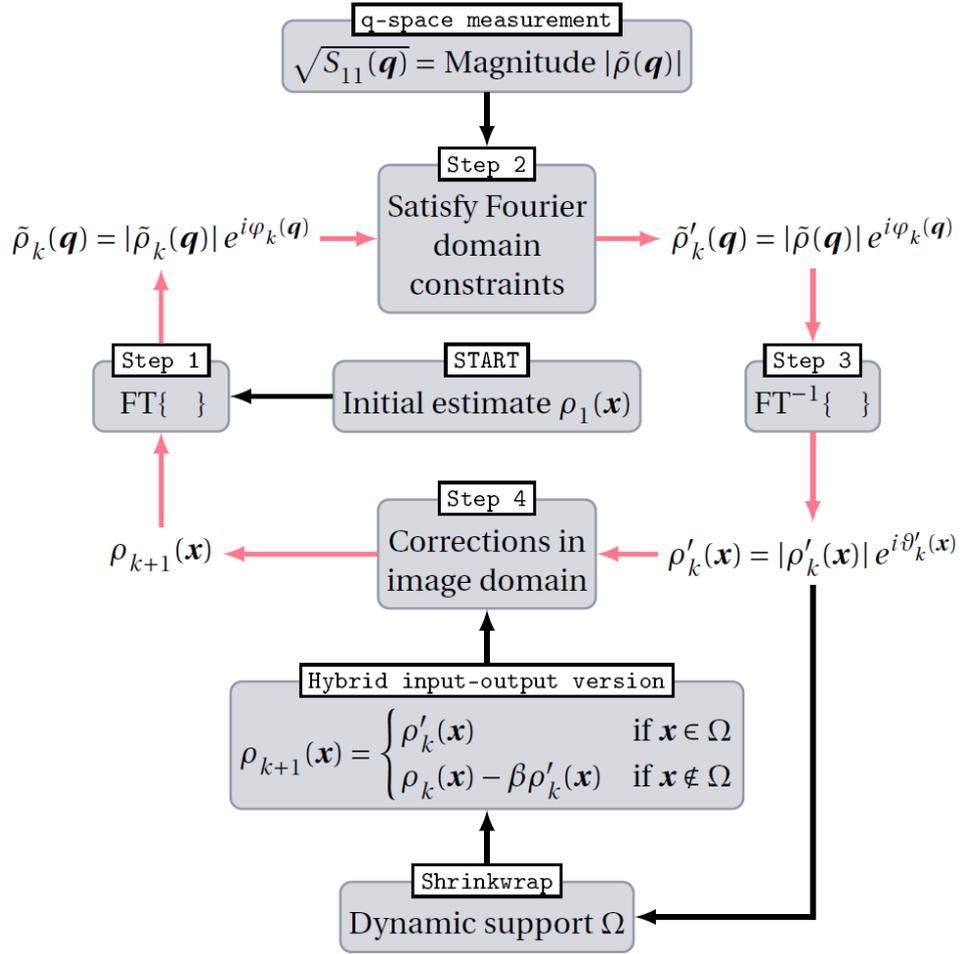

**FIG. 2. (One column, color online)**

Block diagram of the iterative phase retrieval algorithm based on [63]. The algorithm is seeded with a random starting image. Afterwards, the algorithm iterates in four steps as illustrated by the red arrows between Fourier ($q$) and image ($x$) space using corresponding Fourier transforms (step 1 & 3). In step 2, the estimated Fourier magnitude is replaced with the measured magnitude from $q$-space imaging while keeping the estimated phase. In step 4, a support constraint is enforced. In this case the hybrid input-output (HIO) version is depicted.

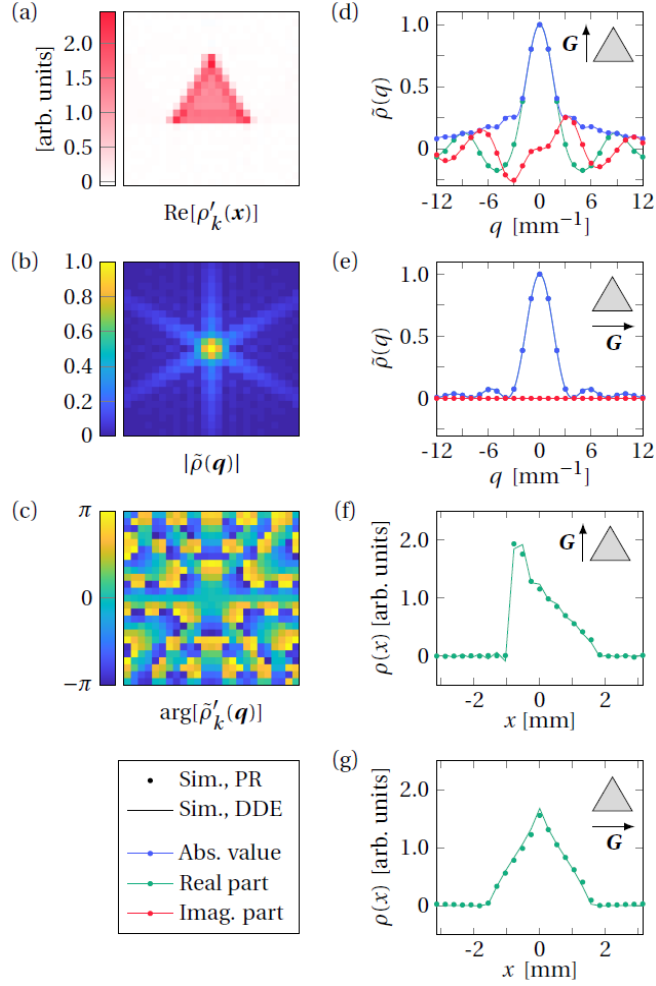

**FIG. 3. (One column, color online)**

Phase retrieval results for diffusion simulations for the equilateral triangle. (a) Final pore image $\rho'_k(x)$. The magnitude of the phase retrieval result in $q$ space (b) is given by the simulated diffraction pattern by construction of the algorithm. The retrieved phase (c) is conjugate symmetric, i.e., $\tilde{\rho}'_k(q) = \tilde{\rho}'^{*}_k(-q)$, as expected. For two gradient directions indicated by the vector $G$, $q$-space profiles (d,e) and pore image projections (f,g) of the phase retrieval result (PR, dots) are compared to the recursive phase reconstruction method using DDE from section II.C (DDE, lines).

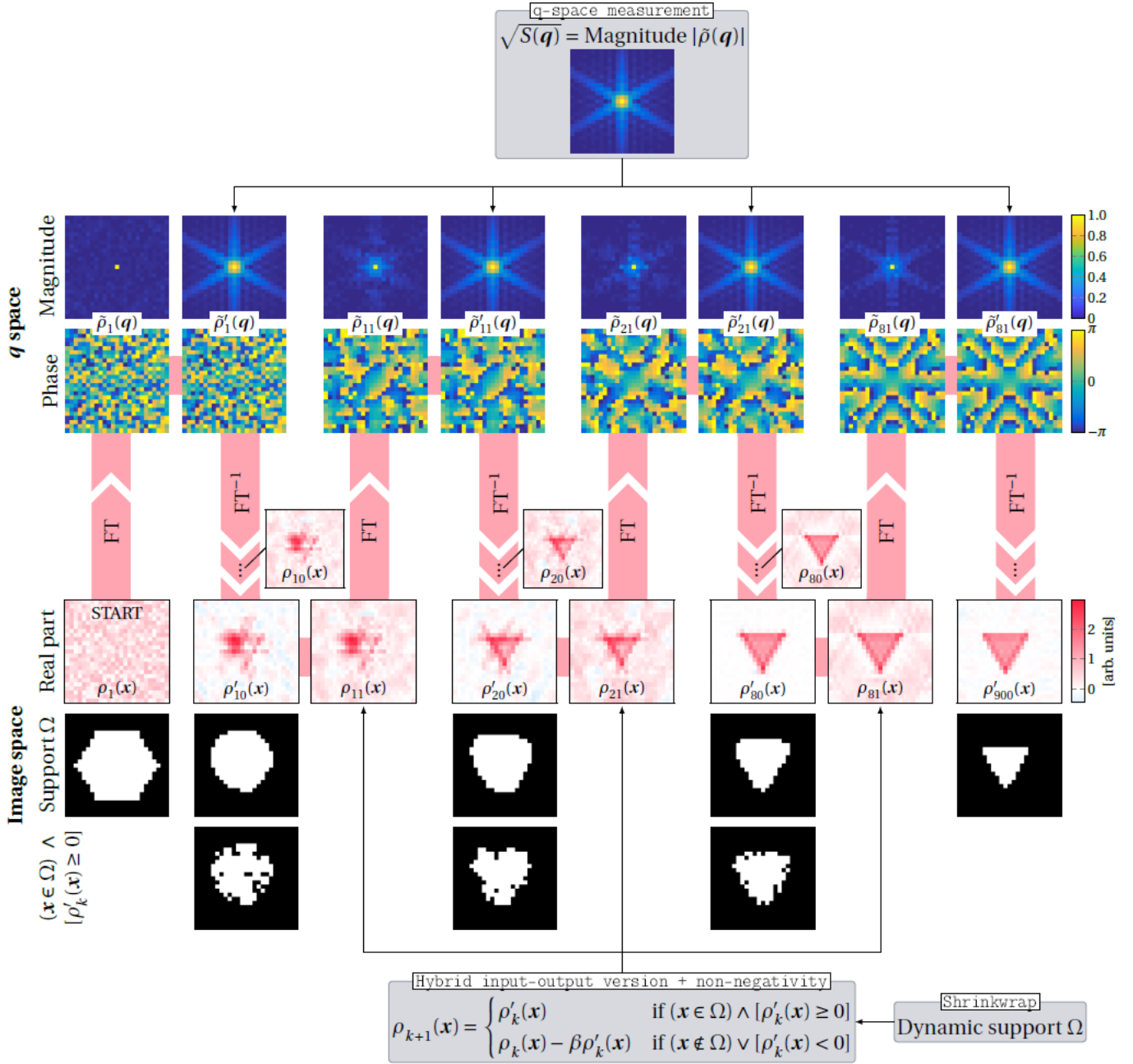

**FIG. 4. (Two column, centered, color online)**

Example on how the HIO algorithm with shrinkwrap extension progresses following the scheme in Fig. 2 plus additional non-negativity constraint. Following the meandering red line, the algorithm starts with a random guess and then cycles back and forth between image and $q$ space until the maximum number of 2000 iterations is reached. Thresholding the autocorrelation function gives the initial support, which shrinks down form the hexagonal shape to the final triangular pore shape with increasing number of iterations $k$. Since only a limited number of iterations could be shown, sequences of three dots were inserted for the skipped iterations. Images of $\mathrm{Re}[\rho'_k(x)]$: Colors for negative (blue) and positive (red) values of are chosen such that the gray levels represent $|\rho'_k(x)|$ in the printed grayscale version.

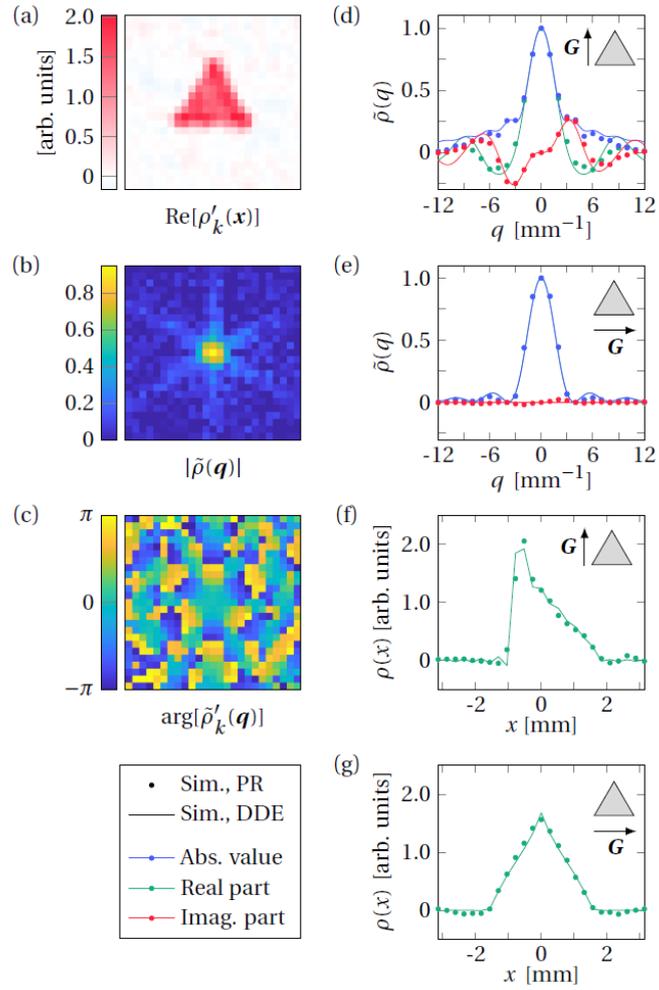

**FIG. 5. (One column, color online)**

Phase retrieval results for diffusion simulations with added noise for the equilateral triangle. Except for that the signal-to-noise ratio of $S_{11}(q)$ was set to 150 at q=0 mm$^{-1}$, the figure is identical to Fig. 3. In (a), colors for negative and positive values of $\mathrm{Re}[\rho'_k(x)]$ are chosen such that in the printed grayscale version the gray levels represent $|\rho'_k(x)|$.

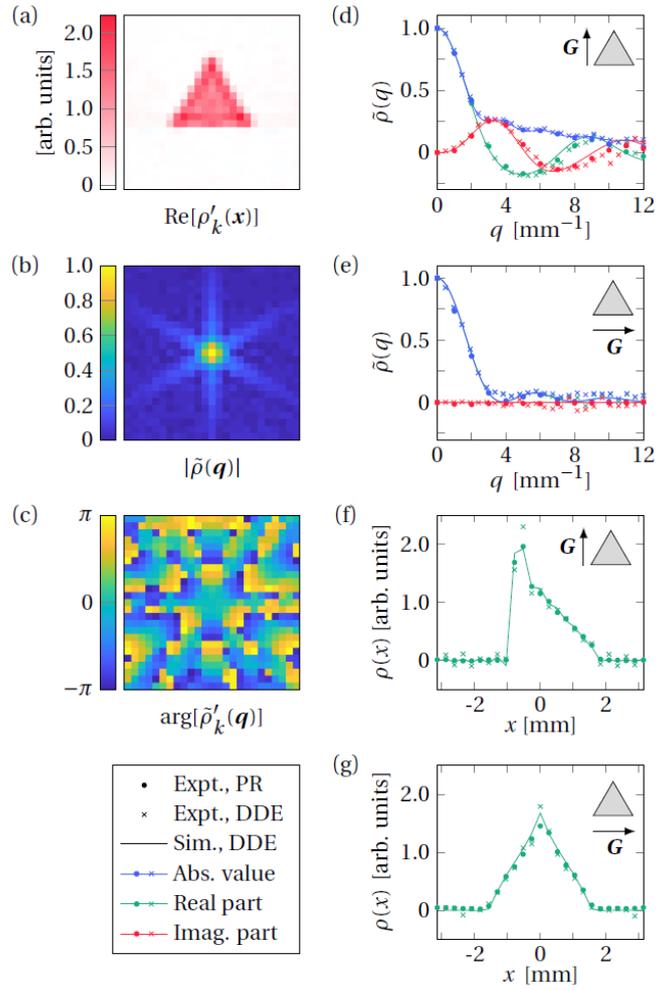

**FIG. 6. (One column, color online)**

Phase retrieval results for phantom experiments with the equilateral triangle. (a) Final pore image. (b) Measured magnitude in $q$ space. (c) Retrieved phase. For two gradient directions indicated by the vector $\mathbf{G}$, $q$-space profiles (d,e) and pore image projections (f,g) of the phase retrieval result (PR, dots) are compared to the recursive phase reconstruction method for measured (DDE, crosses) and simulated DDE signals (DDE, lines). For better visibility, only positive $q$ values are plotted.

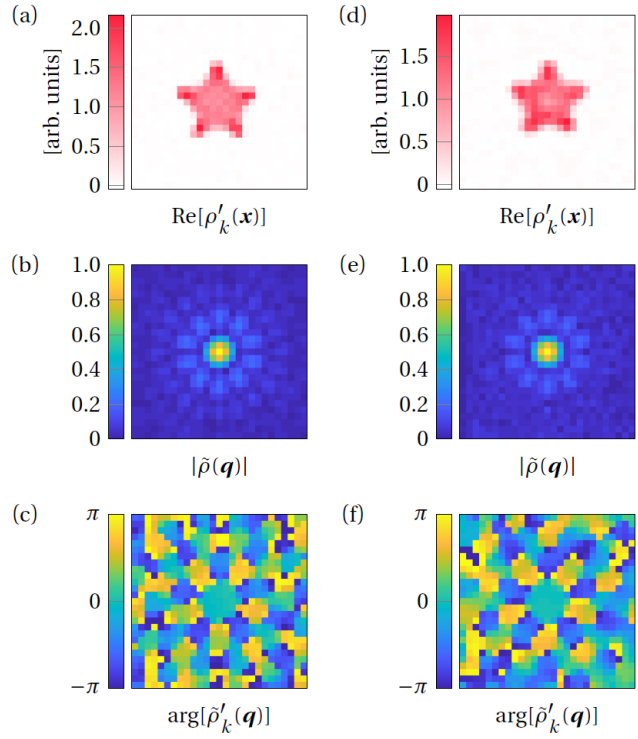

**FIG. 7. (One column, color online)**

Phase retrieval results for phantom experiments with the star-shaped domain. (a) – (c) show the phase retrieval result for simulated $q$-space data of star-shaped pores, which exhibits noise because the Monte Carlo method was used instead of the eigenvalue decomposition approach, and (d) – (f) correspondingly for measured data.

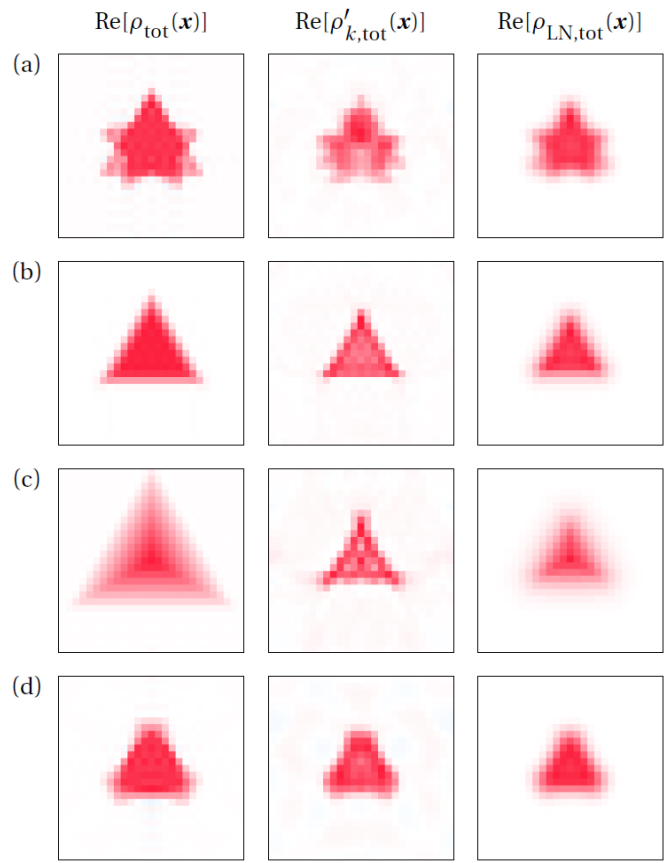

**FIG. 8. (One column, color online)**

Simulations of pore distributions: (a) Shape distribution containing one triangular and one star-shaped pore. (b) Narrow and (c) broad size distribution of triangles. (d) Orientation distribution of triangles. Columns: (1) Average pore image for $\delta \to 0$, $T \to \infty$. (2) Result for the phase retrieval method and for the (3) long-narrow approach using finite gradient pulses.